\documentclass[moor]{informs1}              % for a regular run
\usepackage{natbib}
 \NatBibNumeric
 \def\bibfont{\small}%
 \bibpunct[, ]{[}{]}{,}{n}{}{,}%

\usepackage[colorlinks=true,breaklinks=true,bookmarks=true,urlcolor=blue,
     citecolor=blue,linkcolor=blue,bookmarksopen=false,draft=false]{hyperref}

\def\EMAIL#1{\href{mailto:#1}{#1}}% When hyperref is used, otherwise outcomment 
\TheoremsNumberedThrough     % Preferred (Theorem 1, Lemma 1, Theorem 2)
\EquationsNumberedThrough    % Default: (1), (2), ...
\begin{document}
%%%%%%%%%%%%%%%%

%\setcounter{page}{1}
%\DOI{10.1287/xxxx.0000.0000}%

 \RUNAUTHOR{Kavaler and Smorodinsky}

\RUNTITLE{A Cardinal Comparison of Experts}
%\RUNTITLE{}

% \TITLE{Bundling Information Goods of Decreasing Value}
\TITLE{A Cardinal Comparison of Experts}
%\TITLE{}

\ARTICLEAUTHORS{
\AUTHOR{Itay Kavaler, Rann Smorodinsky}
\AFF{Davidson Faculty of Industrial Engineering and Management, Technion, Haifa 3200003, Israel
\\  \EMAIL{itayk@campus.technion.ac.il} {(IK)}; \EMAIL {rann@ie.technion.ac.il} (RS)}
} % end of the block

\ABSTRACT{%
In various situations, decision makers face experts
that may provide conflicting advice. This advice may be in the form
of probabilistic forecasts over critical future events. We consider
a setting where the two forecasters provide their advice repeatedly
and ask whether the decision maker can learn to compare and rank the
two forecasters based on past performance. We take an axiomatic approach
and propose three natural axioms that a comparison test should comply
with. We propose a test that complies with our axioms. Perhaps, not
surprisingly, this test is closely related to the likelihood ratio
of the two forecasts over the realized sequence of events. More surprisingly,
this test is essentially unique. Furthermore, using results on the
rate of convergence of supermartingales, we show that whenever the
two experts\textquoteright{} advice are sufficiently distinct, the
proposed test will detect the informed expert in any desired degree
of precision in some fixed finite time. % Enter your abstract
}%

%\KEYWORDS{deterministic inventory theory; infinite linear programming duality; 
%  existence of optimal policies; semi-Markov decision process; cyclic schedule}
%\MSCCLASS{Primary: 90B05; secondary: 90C40, 90C90}
%\ORMSCLASS{Primary: Inventory/production: deterministic multi-item;
%  secondary: dynamic programming/optimal control: deterministic 
%  semi-Markov; programming: infinite dimensional}
%\HISTORY{Received November 20, 2003; revised March 8, 2004, and March 26, 2004.}

\MSCCLASS{C11; C70; C73; D83}
%\ORMSCLASS{Primary: ; secondary: }

\maketitle
%%%%%%%%%%%%%%%%%%%%%%%%%%%%%%%%%%%%%%%%%%%%%%%%%%%%%%%%%%%%%%%%%%%%%%

% Samples of sectioning (and labeling) in MOOR.
% NOTE: (1) all section levels end with a period,
%       (2) capitalization is as shown (sentence style, not title style).
%
%\section{Introduction.}\label{intro} %%1.
%\subsection{Duality and the classical EOQ problem.}\label{class-EOQ} %% 1.1.
%\subsection{Outline.}\label{outline1} %% 1.2.
%\subsubsection{Cyclic schedules for the general deterministic SMDP.}
%  \label{cyclic-schedules} %% 1.2.1
%\section{Problem description.}\label{problemdescription} %% 2.

% Text of your paper here
\section{Introduction}\

Consider an individual who repeatedly consults two weather forecasting
websites. It is reasonable to ask what should the individual do when
the two forecasts repeatedly contradict. In what way can the individual
rank the two? Should the individual trust one site and (eventually)
ignore the other? 

The weather example above serves as a metaphor for a plethora of settings
where a decision maker faces conflicting expert advice. Take for example
an elected official who must rely on professional input from civil
servants, a patient who receives prognosis from various doctors or,
more abstractly, a learning algorithm mechanism that uses input from
various sources.

In this paper, we set the stage for defining the notion of a {\em
cardinal comparison test}. The setting we have in mind is a sequential
one. At each stage $t$ two forecasters provide a probability over
some future event (e.g., the occurrence of rain) and then the event
is either realized or its complement is. Before the next day's forecasts
the test must rank the two forecasters. We calibrate these ranks so
they add up to one. One way to think of the rank is a recommendation
for a coin flip to decide which of the two experts' advice should
be taken. 

We pursue a test that complies with the following set of properties
which we consider natural:

\textbf{Anonymity} - A test is {\em anonymous} if it does not depend
on the identity of the experts but only on their forecasts.

\textbf{Error-free} - A test is {\em error-free} if from their
perspective, each of the experts cannot entertain the thought that
the other expert will be overwhelmingly preferred (i.e., he assigns relatively lower
probability). Another way to think about a notion of an error-free
test is to assume that one of the experts has the correct model. In
such a case, the test will probably not point at the second expert as
the superior one.

\textbf{Reasonable} - Let us consider an event, $A$, that has positive
probability according to the first expert but relatively lower probability according
to the second. Conditional on the occurrence of event $A$, a {\em
reasonable} test must assign positive probability to the first expert
being better informed than the second.

One thing to emphasize about the cardinal comparison test we pursue and the
related properties is that they are not designed to evaluate whether
either of the two forecasters is correct in some objective sense.
They are only designed to compare the two. To make this point, assume
that Nature follows a fair coin for deciding on rain and one forecaster
insists on forecasting rain with probability 60\% while the other
insists on 10\%. While both are wrong, a cardinal comparison test should somehow
gravitate towards the former one as being better. 

There is a large body of literature on expert testing that studies
the question of whether a self-proclaimed expert is a true expert
or a charlatan (see Section 1.2 for more details) and many of the
results point to the difficulty or impossibility of designing such
tests that are immune to strategic forecasters. 

A comparison test may often be a more natural question than the one
on whether the forecaster is correct. Indeed, when a decision maker
must act, then she must choose which of the experts to follow. In the
case of a single expert, the dismissal of that expert leaves the decision
maker working with her own unsubstantiated beliefs, which may lead
to an even worse outcome. In case a decision maker faces two forecasters
with conflicting input, she may choose to somehow aggregate the two
instead of dismissing one or the other. We discuss this alternative
line of research in Section 1.2.

\subsection{Results}\

Given an ordered pair of forecasters, $f$ and $g$, at any finite
time $t$, we consider the corresponding likelihood ratio of the actual
outcome and calibrate it so that it and its inverse add up to one.
We call this the finite derivative test at time $t$. We prove that
this test is anonymous, error-free and reasonable. Furthermore, modulo
an equivalence relation, it is unique. In fact, for any test that
differs from the aforementioned construction and which is anonymous
and reasonable, there exist two forecasters which render the test
not error-free.

Moreover, our constructed test perfectly identifies the correct forecaster
whenever the two measures induced by the forecasters are mutually
singular with respect to each other. Requiring the test to identify
the correct expert when the measures are not mutually singular is
shown to be impossible.

A test could potentially take a long while until it converges to a
verdict on the better expert. We show that the proposed comparison
test converges fast and uniformly. In fact, when disregarding the
stages at which the two experts provide similar forecasts, then with
high probability the correct verdict will emerge in finite time that
is independent of the underlying probabilities.

One can ask whether {\em ideal} tests can exist, that is, tests
that always rank the correct forecaster higher regardless of what
forecasting strategies other experts might submit. Unfortunately,
this turns out to be impossible, as we discuss in Appendix \ref{Appendinx sec:On-ideal-tests}.
Since an ideal test does not exist, it is natural to explore the ideality
of a test over a limited class of data-generating processes. We provide
a full characterization for the existence of ideal tests over sets
by showing that an ideal test with respect to a set $A$ exists  if
and only if, $A$ is pairwise mutually singular.

\subsection{\label{subsec:Related-Literature}Related literature}\

Single expert testing. A substantial part of the literature on expert
testing focuses on the single expert setting. This literature dates
back to the seminal paper of \citet*{Dawid-1982}, who proposes the
calibration test as a scheme to evaluate the validity of weather forecasters.
Dawid asserts that a test must not fail a true expert. \citet{Foster-Vohra-1998}
show how a charlatan, who has no knowledge of the weather, can produce
forecasts which are always calibrated. The basic ingredient that allows
the charlatan to fool the test is the use of random forecasts. \citet*{Lehrer-2001}
and \citet*{Sandroni-and-Smorodinsky-and-Vohra-2003} extend this
observation to a broader class of calibration-like tests. Finally,
\citet*{Sandroni-2003} shows that there exists no error-free test
that is immune to such random charlatans (see also extensions of Sandroni's
result in \citet{Shmaya-2008} and \citet{Olszewski-Sandroni-2008}).

To circumvent the negative results, various authors suggest limiting
the set of models for which the test must be error-free (e.g., \citet*{Al-Najjar-2010}
and \citet*{Pomatto-2016}), or limiting the computational power associated
with the charlatan (e.g., \citet{Fortnow-Vohra-2009}) or replacing
measure theoretic implausibility with topological implausibility by
resorting to the notion of {}category one sets{} (e.g., \citet*{Dekel-Feinberg}).

Multiple expert testing. Comparing performance of two (or more) experts
gained very little attention in the literature. Apart from our previous
work, \citet{Kavaler-Smorodinsky-2019}, we are only familiar with
\citet*{Al-Najjar-2008}. That paper proposes a test based on the
likelihood ratio for comparing two experts. They show that if one
expert knows the true process whereas the other is uninformed, then
one of the following must occur: either, the test correctly identifies
the informed expert, or the forecasts made by the uninformed expert
are close to those made by the informed one. It turns out that the
test they propose is anonymous and reasonable but is not error-free
(please refer to Section \ref{sec:Independent-axioms} for the formal
definition).

Another approach was suggested by \citet*{Feinberg-Stewart-2008},
who study an infinite horizon model of testing multiple experts, using
a cross-calibration test. In their test, $N$ experts are tested simultaneously;
each expert is tested according to a calibration restricted to dates
where not only does the expert have a fixed forecast but the other
experts also have a fixed forecast, possibly with different values.
That is to say, where the calibration test checks the empirical frequency
of observed outcomes conditional on each forecast, the cross-calibration
test checks the empirical frequency of observed outcomes conditional
on each profile of forecasts (please refer to Appendix \ref{sec:Appendix B The-Cross-Calibration}
for the formal definition).

They showed that if an expert predicts according to the data-generating
process, the expert is guaranteed to pass the cross-calibration test
with probability 1, no matter what strategies the other experts use.
In addition, they prove that in the presence of an informed expert,
the subset of data-generating processes under which an ignorant expert
(a charlatan) will pass the cross-calibration test with positive probability,
is topologically ``small''. 

In a previous paper, \citet{Kavaler-Smorodinsky-2019}, we construct
a comparison test over the infinite horizon. In that paper, the test
outputs one verdict at the end of all times which is in one of three
forms-{}-{}-it points to either one of the forecasters as advantageous
or it is indecisive. The main result in that paper was the identification
of an essentially unique infinite-horizon, ordinal test that adheres
with some natural properties. The properties studied in the current
paper (as well as the associated terminology) are inspired by the
ones studied in \citet{Kavaler-Smorodinsky-2019}. The test we identify
is based on the likelihood ratio. Interestingly, the tests identified
in \citet{Al-Najjar-2008} and that identified by \citet{Pomatto-2016}
for testable paradigms are also based on the likelihood ratio.

An alternative approach to that of comparing and ranking experts is
that of aggregating forecasts by a non-Bayesian aggregator. For aggregation
schemes that do well in a single stage setting, see \citet{Arieli-and-Babichenko-and-Smorodinsky-2018},
as well as \citet{Levy-2018}, and \citet{Razin-2018}; for schemes
that work well in a repeated setting and produce small regret, see
the rich literature in machine learning surveyed in \citet{Cesa-Bianchi:2006}.
\vspace{0cm}

\section{Model\label{sec:Model}}\

At the beginning of each period $t=1,2,\ldots$ an outcome, $\omega_{t}$,
drawn randomly by Nature from the set $\Omega=\{0,1\},$ is realized.\footnote{{}For expository reasons, we restrict attention to a binary set $\Omega=\{0,1\}$.
The results extend to any finite set.} A {\em realization} is an infinite sequence of outcomes, $\omega:=\{\omega_{1},\omega_{2},\ldots\}\in\Omega^{\infty}$.
We denote by $\omega^{t}:=\{\omega_{1},\omega_{2},\ldots,\omega_{t}\}$
to be the prefix of length $t$ of $\omega$ (sometimes referred to
as the partial history of outcomes up to period $t$) and use the
convention that $\omega^{0}:=\emptyset$. At the risk of abusing notation,
we will also use $\omega^{t}$ to denote the cylinder set $\{\hat{\omega}\in\Omega^{\infty}:\hat{\omega}^{t}=\omega^{t}\}.$
In other words, $\omega^{t}$ will also denote the set of realizations
which share a common prefix of length $t$. For any $t$ we denote
by ${\cal G}_{t}$ the $\sigma$-algebra on $\Omega^{\infty}$ generated
by the cylinder sets $\omega^{t}$ and let 
${\cal G}_{\infty}:=\sigma(\stackrel[t=0]{\infty}{\bigcup}{\cal G}_{t})$
denote the smallest $\sigma$-algebra which consists of all cylinders
(also known as the Borel $\sigma$-algebra). Let $\Delta(\Omega^{\infty})$
be the set of all probability measures defined over the measurable
space $(\Omega^{\infty},{\cal G}_{\infty})$.

Before $\omega_{t}$  is realized, two self-proclaimed experts (sometimes
referred to as forecasters) simultaneously announce their forecast
in the form of a probability distribution over $\Omega$. Let $(\Omega\times\Delta(\Omega)\times\Delta(\Omega))^{t}$
be the set of all sequences composed of realizations and pairs of
forecasts made up to time $t$ and let $\underset{t\geq0}{\bigcup}(\Omega\times\Delta(\Omega)\times\Delta(\Omega))^{t}$
be the set of all such infinite sequences.

A (pure) forecasting strategy $f$ is a function that maps finite
histories to a probability distribution over $\Omega$. Formally,
$f:\underset{t\geq0}{\bigcup}(\Omega\times\Delta(\Omega)\times\Delta(\Omega))^{t}\longrightarrow\Delta(\Omega).$
Note that each forecast provided by one expert may depend, inter alia,
on those provided by the other expert in previous stages. Let $F$
denote the set of all forecasting strategies.

A probability measure $P\in\Delta(\Omega^{\infty})$ naturally induces
a (set of) corresponding forecasting strategy, denoted $f_{P}$, that
satisfies for any $\omega\in\Omega^{\infty}$ and any $t$ such that
$P(\omega^{t})>0$ 
\[
f_{P}(\omega^{t},\cdot,\cdot)(\omega_{t+1})=P(\omega_{t+1}|\omega^{t}).
\]

\noindent Thus, the forecasting strategy $f_{P}$ derives its forecasts
from the original measure, $P$, via Bayes rule. Note that this does
not restrict the forecast of $f_{P}$ over cylinders, $\omega^{t}$,
for which $P(\omega^{t})=0$. \footnote{An expert who uses $f_{P}$ to derive the correct predictions is referred
to as informed, whereas an expert who concocts predictions strategically
to pass the test without any knowledge on $P$ is referred to as uninformed.}

In the other direction, a realization $\omega,$ and an ordered pair
of forecasting strategies, $\vec{f}:=(f,g),$ induce a unique play
path, $(\omega,\vec{f})\in(\varOmega\times\Delta(\Omega)\times\Delta(\Omega))^{\infty},$
 where the corresponding $t$ - history is denoted by $(\omega,\vec{f})^{t}\in(\Omega\times\Delta(\Omega)\times\Delta(\Omega))^{t}$
started at the Null history, $(\omega,\vec{f})^{0}:=\emptyset,$ which
in turn induce a  pair of probability measures, denoted for simplicity
by $(f,g)$, over $\Omega^{\infty}$, as follows: 
\[
f(\omega^{t})=\stackrel[n=1]{t}{\prod}f((\omega,\vec{f})^{n-1})[\omega_{n}],\;g(\omega^{t})=\stackrel[n=1]{t}{\prod}g((\omega,\vec{f})^{n-1})[\omega_{n}].
\]
By Kolomogorov\textquoteright s extension theorem, the above is sufficient
in order to derive the whole measure. Observe that a pair of forecasting
strategies induces a pair of probability measures, whereas each single
forecasting strategy does not induce a single measure due to the dependency
between the two forecasters.

\subsection{{\label{subsec:A-finite-comparison} A cardinal comparison
test}}\

At each stage $t$ a third party (the `tester') who observes the
forecasts and outcomes compares the performance of both forecasters
and decides who she thinks is better. Formally, 

\vspace{0cm}

\begin{definition}
\label{Def a cardinal comparison test}A {\em cardinal comparison test}
is a sequence $T:=(T_{t})_{t>0}$, where

\noindent $T_{t}:(\Omega\times\Delta(\Omega)\times\Delta(\Omega))^{\infty}\longrightarrow[0,1]$
is ${\cal {\cal G}}_{t}-$measurable for all $t>0$.
\end{definition}
\vspace{0cm}

In other words, for any $t$ and any realization $\omega$ and any
sequence of forecasts $\vec{f}$, the tester, conditional on a $t$
- history, announces her level of confidence that the first forecaster
(the one using $f$) is better than the second one (we will interchangeably
refer to this as his propensity that $f$ is superior to g).\footnote{It should be emphasized that the results in this paper hold even for
the general case for which definition \ref{Def a cardinal comparison test}
is extended such that a tester may condition his one step ahead decisions
on his own past decisions. Formally, whenever $T$ has the form $T_{t}:\ (\Omega\times\Delta(\Omega)\times\Delta(\Omega)\times\{f,g\})^{\infty}\longrightarrow[0,1].$} Note that announcing $0.5$ means that both are equally capable (this
should not be confused with the statement that they are both capable
or both incapable). Whenever $T_{t}(\omega,\vec{f})=1$ (respectively,
$0$) the tester is confident that $f$ outperforms $g$ (respectively,
$g$ outperforms $f$).

\vspace{0cm}

\begin{definition}
\label{Def anonymity}$T$ is called {\em anonymous} if for all $\omega\in\Omega^{\infty},t>0$
and for all $f,g\in F,$
\[
T_{t}(\omega,f,g)=1-T_{t}(\omega,g,f).
\]
\end{definition}
\noindent In other words, the test's propensity at each period should
not depend on the expert's identity. Note that whenever $f=g$ an
anonymous test $T$ must output a propensity of $0.5$ for all $\omega\in\Omega^{\infty},\ t>0$. 

For a given test $T,$ an ordered pair of forecasting strategies $\vec{f}=(f,g),$
and a realization $\omega$, we denote by $T(\omega,\vec{f})\:= limT_{t}(\omega,\vec{f})$
whenever the limit exists. For $\epsilon\in(0,1)$, let $L_{T,\epsilon}^{\vec{f}}:=\{\omega\colon\ T(\omega,\vec{f})>\epsilon\}$
be the set of realizations for which the limit of $T$ exists and
from some time on assigns a propensity larger than $\epsilon$ to
$f$ (similarly we denote$\ R_{T,\epsilon}^{\vec{f}}:=\{\omega\colon\ T(\omega,\vec{f})<\epsilon\}).$
Notice that the following is a straightforward observation derived
from Definition \ref{Def anonymity}. If $T$ is an anonymous test,
then $\omega\in R_{T,\epsilon}^{(f,g)}$ if and only if $\omega\in L_{T,1-\epsilon}^{(g,f)}$;
we use the last for some of our proofs.

When $\omega$ is in $L_{T,\epsilon}^{\vec{f}}$ and $\epsilon>0.5$,
the test eventually assigns a higher propensity to $f$ than to $g$.
On the other hand, for $\epsilon<0.5$, the test assigns a higher
propensity to $g$ whenever $\omega$ is in $R_{T,\epsilon}^{\vec{f}}.$
Thus, we will typically focus on the sets $L_{T,\epsilon}^{\vec{f}}$
with $\epsilon>0.5$ and on the sets $R_{T,\epsilon}^{\vec{f}}$ for
$\epsilon<0.5$.

\vspace{0cm}

\subsection{{\label{subsec:Desirable-Properties}Desirable Properties}}\

In this section, we introduce a set of axioms we deem desirable for
a cardinal comparison test. Our first property asserts that any set
that is contained in $R_{T,\epsilon}^{\vec{f}}$ must not be assigned
a high probability according to $f$ in comparison with the probability
assigned by $g$. In particular, the ratio of these probabilities
must be bounded by $\frac{\epsilon}{1-\epsilon}.$
\begin{definition}
\label{DEF error-free}$T$ is {\em error-free} if for all $\vec{f}:=(f,g)\in F\times F,$
for all $\epsilon\in(0,\frac{1}{2})$ and for all measurable set $A$

\begin{equation}
f(A\cap R_{T,\epsilon}^{\vec{f}})\leq(\frac{\epsilon}{1-\epsilon})g(A\cap R_{T,\epsilon}^{\vec{f}})\label{eq: error free condition}
\end{equation}

\noindent (Similarly, $g(A\cap L_{T,\epsilon}^{\vec{f}})\leq(\frac{1-\epsilon}{\epsilon})f(A\cap L_{T,\epsilon}^{\vec{f}})$
for $\epsilon\in(\frac{1}{2},1)$).
\end{definition}
Note, in particular, as $\epsilon$ approaches $0$, the set $R_{T,\epsilon}^{\vec{f}}$
captures the paths where $g$ is clearly deemed better than $f$ and
so the property of error-freeness implies that although $g$ may assign
a subset of $R_{T,\epsilon}^{\vec{f}}$ a positive probability, it
must be the case that $f$ assigns it near-zero probability. On the
other hand, whenever $\epsilon$ approaches $0.5$, the corresponding
ratio approaches $1$ and so error-freeness requires that $f$ assigns
that event a probability no greater than $g$.

In particular, each forecaster must believe that a test cannot point
out the other forecaster as correct. From his perspective, he is either
preferred or the test is indecisive.

\vspace{0cm}

Consider a set of realizations assigned positive probability by one
forecaster whereas his colleague assigns it a relatively lower probability.
We shall call a test `reasonable' if the former forecaster assigns
a positive probability to the event that the test will eventually
provide a high propensity to her. Formally:

\vspace{0cm}

\begin{definition}
\label{DEF resonable} $T$ is {\em reasonable} if for
all $\vec{f}\in F\times F,$ for all $\epsilon\in(0,\frac{1}{2})$
and for all measurable set $A,$ 

\begin{equation}
g(A)>0\ and\ f(A)<(\frac{\epsilon}{1-\epsilon})g(A)\Longrightarrow\text{ }g(A\cap R_{T,\epsilon}^{\vec{f}})>0.\label{eq:reasonableness condition}
\end{equation}

\noindent (Similarly, $f(A)>0\ and\ g(A)<(\frac{1-\epsilon}{\epsilon})f(A)\Longrightarrow f(A\cap L_{T,\epsilon}^{\vec{f}})>0$
for $\epsilon\in(\frac{1}{2},1)$).
\end{definition}
\noindent It should be emphasized that reasonableness and error-freeness
are not related notions; examples that these properties are independent
will be discussed in Section \ref{sec:Independent-axioms}.

\vspace{0cm}

\begin{remark}
One could propose to replace error-freeness with a stronger and more
appealing property in which a test points out the better informed
expert with probability one. Informally, we would like to consider
tests that have the following property $f(T(\omega,\vec{f})=1)=1$
whenever $f\not=g$. However, there could be pairs of forecasters
that are not equal but induce the same probability distribution. In
appendix \ref{Appendinx sec:On-ideal-tests}, we formalize this and
refer to tests that satisfy this stronger requirement as an {\em ideal}.
We, furthermore show, as the name suggests, that such tests essentially
do not exist.
\end{remark}\

\
\section{{An error-free and reasonable test}}\

We now turn to propose an anonymous cardinal comparison test that is error-free
and reasonable. For any pair of forecasters, $\vec{f}:=(f,g)\in F\times F,\ \omega\in\Omega^{\infty},\ t\geq0,$
the {\em finite derivative test}, ${\cal D},$ is defined as follows:
\[
{\cal D}_{t+1}(\omega,\vec{f})=\begin{cases}
\begin{array}{l}
\frac{f(\omega^{t})}{f(\omega^{t})+g(\omega^{t})},\\
\frac{1}{2},
\end{array} & \begin{array}{l}
g(\omega^{t})>0\ or\ f(\omega^{t})>0\\
other.
\end{array}\end{cases}
\]
It should be noted that the ratio between ${\cal D}_{t+1}(\omega,\vec{f})$,
the rank associated with the forecast $f$ and $1-{\cal D}_{t+1}(\omega,\vec{f})$,
the rank associated with the forecast $g,$ equals the likelihood
ratio between the two forecasters. Clearly, ${\cal D}$ is anonymous.
We turn to show that it is reasonable and error-free. Before doing
so, some preliminaries are required.\footnote{Notice that ${\cal D}$ is unaffected by the so-called \textquotedblleft counterfactual\textquotedblright{}
predictions. These predictions are referred to events which may not occur.
On the contrary, the outcome of ${\cal D}$ depends only on predictions
which were made along the realized play path.}
\begin{lemma}
\textup{\label{Lemma 1}Let $\vec{f}:=(f,g)$. Then the limit of ${\cal D}_{t}(\cdot,\vec{f})$
exists and is finite  $f-a.s.$}
\end{lemma}

\proof{Proof.}\ 
For $\omega\in\Omega^{\infty}$ where $f(\omega^{t})>0$ define the
likelihood ratio between the two forecasters at time $t$ as 
\[
D_{f}^{t}g(\omega)=\stackrel[n=1]{t}{\prod}\frac{g((\omega,\vec{f})^{n-1})[\omega_{n}]}{f((\omega,\vec{f})^{n-1})[\omega_{n}]},
\]
\noindent and observe that ${\cal D}_{t+1}(\omega,\vec{f})=\frac{1}{1+D_{f}^{t}g(\omega)}.$\footnote{If $f((\omega,\vec{f})^{n-1})[\omega_{n}]=0$ for some $n$, we set
$D_{f}^{t}g(\omega)=\infty$ for all $t\geq n$.} Applying Lemma 2 from \citet{Kavaler-Smorodinsky-2019}, we know
that the limit of $D_{f}^{t}g,$ denoted $D_{f}g,$ exists and is
finite $f-a.s$. It readily follows that ${\cal D}(\omega,\vec{f}):=\frac{1}{1+D_{f}g(\omega)}:= lim{\cal D}_{t}(\omega,\vec{f})$
exists and is finite $f-a.s.$\Halmos
\endproof

Now that we have established the existence and the finiteness of the
test ${\cal D}$, let us prove that it complies with the two central
properties for cardinal comparison tests: 
\begin{proposition}
\textup{\label{Prop 2: D is  error-free}${\cal D}$ is {\em error-free}.}
\end{proposition}
\proof{Proof.}\
Let $\vec{f}:=(f,g)\in F\times F,\ \epsilon\in(0,\frac{1}{2}),$ and
a measurable set $A.$ From Lemma \ref{Lemma 1}, the limit of ${\cal D}_{t}(\cdot,\vec{f})$ exists and is finite $f-a.s.$ Hence,
\[
A\cap R_{{\cal D},\epsilon}^{\vec{f}}=\{\omega\in A\colon\ \frac{1}{1+D_{f}g(\omega)}\in R_{{\cal D},\epsilon}^{\vec{f}}\}\subset\{\omega\colon\ D_{f}g(\omega)\geq\frac{1-\epsilon}{\epsilon}\}.
\]
\noindent Thus, applying \citet{Kavaler-Smorodinsky-2019}, Lemma
2, part b, we obtain
\[
f(A\cap R_{{\cal D},\epsilon}^{\vec{f}})\leq(\frac{\epsilon}{1-\epsilon})g(A\cap R_{{\cal D},\epsilon}^{\vec{f}}).
\]
\noindent Similarly, by applying \citet{Kavaler-Smorodinsky-2019},
Lemma 2, part a, we show that $g(A\cap L_{{\cal D},1-\epsilon}^{\vec{f}})\leq(\frac{\epsilon}{1-\epsilon})f(A\cap L_{{\cal D},1-\epsilon}^{\vec{f}}),$
and ${\cal D}$ is error-free.\Halmos
\endproof 
\begin{proposition}
\textup{\label{Prop 3 : D is  reasonable}${\cal D}$ is {\em reasonable}.}
\end{proposition}
\proof{Proof.}\
Let $\vec{f}\in F\times F,\ \epsilon\in(0,\frac{1}{2}),$ and a measurable
set $A,$ and suppose (w.l.o.g.) that

\begin{equation}
g(A)>0\ and\ f(A)<(\frac{\epsilon}{1-\epsilon})g(A).\label{eq: reasonable condition for D}
\end{equation}

Denote $A_{1}:=(A\cap L_{{\cal D},\epsilon}^{\vec{f}})\cup(A\cap\{\omega\colon\ {\cal D}(\omega,\vec{f})=\epsilon\}),\ A_{2}:=(A\cap R_{{\cal D},\epsilon}^{\vec{f}})$
and observe that $A=A_{1}\cup A_{2}.$ Assume by contradiction that
$f(A_{2})=0$ and notice that by the construction, 
\[
A_{1}\subset\{\omega\colon\ lim\cal {D}_{t}(\omega,\vec{f})=\frac{1}{1+D_{f}g(\omega)}\geq\epsilon\}.
\]
\noindent Thus, applying \citet{Kavaler-Smorodinsky-2019}, Lemma
2, part a, together with $f(A)=f(A_{1})$, we obtain that $f(A)\geq(\frac{\epsilon}{1-\epsilon})g(A)$
which contradicts \eqref{eq: reasonable condition for D} and hence
$f(A_{2})>0.$ By similar consideration we show that $f(A\cap L_{{\cal D},\epsilon}^{\vec{f}})>0$ whenever 
$f(A)>0\  and\  g(A)<(\frac{1-\epsilon}{\epsilon})f(A)$ for $\epsilon\in({1\over2},1)$, 
and therefore ${\cal D}$ is reasonable. \footnote{\noindent In fact we show a stronger result: since $f$ is monotone
and $R_{{\cal D},\epsilon}^{\vec{f}}=\underset{\bar{\epsilon}\in\mathbb{Q\cap}(0,\epsilon]}{\bigcup}R_{{\cal D},\bar{\epsilon}}^{\vec{f}}$
it follows that, conditional on $f(A_{2})>0$ there exists $\text{\ensuremath{\bar{\epsilon}<\epsilon} such that }g(A\cap R_{{\cal D},\bar{\epsilon}}^{\vec{f}})>0$.}\Halmos
\endproof
Propositions \ref{Prop 2: D is  error-free} and \ref{Prop 3 : D is  reasonable}
jointly prove our first main theorem:
\begin{theorem}
\textup{\label{Thorem:  D is error-free and reasonable}${\cal D}$
is an anonymous, reasonable and error-free test.}
\end{theorem}
We now turn to show that the finite derivative test is essentially
the unique anonymous cardinal comparison test that is reasonable and error-free.

\section{{Uniqueness}}\

Although there may be other error-free and reasonable cardinal comparison tests,
they are essentially equivalent to the finite derivative test. To
motivate this idea, consider the following example.
\begin{example}

Consider the realization $\tilde{\omega}:=(1,1,1,,,),$ and two forecasters $\tilde{f}$ and $\tilde{g}$, both using a coin to make predictions. $\tilde{f}$ uses a fair coin whereas $\tilde{g}$ uses a biased coin with probability one for the outcome to be $1$. Let $\overrightarrow{h_{1}^{t}}$ be the history of length $t$ induced by $(\text{\ensuremath{\tilde{\omega}}},\tilde{f},\tilde{g})$ and let $\overleftarrow{h_{1}^{t}}$ be the one induced by $(\tilde{\omega},\tilde{g},\tilde{f})$. Let $c>1$ and consider the following test:
\[
T_{t}(\omega,\vec{f})=\begin{cases}
\begin{array}{l}
{\cal D}_{t}(\omega,\vec{f}),\\
\frac{1}{1+c\cdot D_{f}^{t}g(\omega)},\\
1-\frac{1}{1+c\cdot D_{g}^{t}f(\omega)},
\end{array} & \begin{array}{l}
other\\
(\omega,\vec{f})^{t}=\overrightarrow{h_{1}^{t}}\\
(\omega,\vec{f})^{t}=\overleftarrow{h_{1}^{t}}.
\end{array}\end{cases}
\]
\noindent Hence, the propensities of $T$ differ from those provided
by ${\cal D}$ only along the play paths $\overrightarrow{h_{1}},\overleftarrow{h_{1}},$
in which case the limit of $T$ converges slower to $1,0,$ respectively,
than ${\cal D}.$
\end{example}
\begin{claim}
$T$ is an anonymous error-free and a reasonable test.
\end{claim}

\proof{Proof.}\
Let $\vec{f}:=(f,g)\in F\times F,\ \epsilon\in(0,\frac{1}{2})$ and a measurable set $A$. Recall that $\vec{f}$ and $\tilde{\omega}$ induce a unique play path, $(\tilde{\omega},\vec{f})$. Thus, if $\tilde{\omega}\notin A\cap R_{T,\epsilon}^{\vec{f}}$ or $\tilde{\omega}\in A\cap R_{T,\epsilon}^{\vec{f}}$ and $(\tilde{\omega},\vec{f})\neq\overleftarrow{h_{1}}$, then the construction yields $A\cap R_{T,\epsilon}^{\vec{f}}=A\cap R_{{\cal D},\epsilon}^{\vec{f}}$. In addition, note that $(\tilde{\omega},\vec{f})=\overrightarrow{h_{1}}$ implies that $T_{t}(\omega,\vec{f})\longrightarrow1$, in which case $\tilde{\omega}\notin R_{T,\epsilon}^{\vec{f}}$. If, on the other hand, $\tilde{\omega}\in A\cap R_{T,\epsilon}^{\vec{f}}$ and $(\tilde{\omega},\vec{f})=\overleftarrow{h_{1}}$ then $T_{t}(\omega,\vec{f})\longrightarrow0$ as $cD_{g}^{t}f(\tilde{\omega})\longrightarrow0$. In which case $f(\tilde{\omega})=\tilde{f}(\tilde{\omega})=0$. Since by Propositions \ref{Prop 2: D is  error-free} ${\cal D}$ is error-free the following is obtain

\[f(A\cap R_{T,\epsilon}^{\vec{f}})=f(A\setminus\tilde{\omega}\cap R_{T,\epsilon}^{\vec{f}})=f(A\setminus\tilde{\omega}\cap R_{{\cal D},\epsilon}^{\vec{f}})\leq(\frac{\epsilon}{1-\epsilon})g(A\setminus\tilde{\omega}\cap R_{{\cal D},\epsilon}^{\vec{f}})\leq(\frac{\epsilon}{1-\epsilon})g(A\cap R_{T,\epsilon}^{\vec{f}}).
\]

The case for which $\epsilon\in(\frac{1}{2},1)$ is analogous and hence omitted. We therefore conclude that T is error-free.

To see why T is reasonable assume that $g(A)>0$ and $f(A)\leq(\frac{\epsilon}{1-\epsilon})g(A)$. Similar consideration shows that either $\tilde{\omega}\in A\cap R_{T,\epsilon}^{\vec{f}}$ and $(\tilde{\omega},\vec{f})=\overleftarrow{h_{1}}$, in which case $g(A\cap R_{T,\epsilon}^{\vec{f}})=\tilde{g}(\tilde{\omega})=1$, or $g(A\cap R_{T,\epsilon}^{\vec{f}})=g(A\cap R_{{\cal D},\epsilon}^{\vec{f}})>0$ where the most-right inequality holds since by proposition \ref{Prop 3 : D is  reasonable}  ${\cal D}$ is a reasonable test. The proof for $\epsilon\in(\frac{1}{2},1)$ is analogous. Finally, by construction, the anonymity of ${\cal D}$ implies the anonymity of $T$.\Halmos
\endproof

To capture the concept of equivalence, we introduce the following
equivalence relation over tests;
\begin{definition}
\label{Def: equivalence tests}Let $\vec{f}:=(f,g)\in F\times F.$
We say that {\em $\text{T}\sim_{\vec{f}}\hat{T}$ } if
\[
f(\{\omega:T(\omega,\vec{f})\ensuremath{\neq\hat{T}(\omega,\vec{f})\})}=g(\{\omega:T(\omega,\vec{f})\ensuremath{\neq\hat{T}(\omega,\vec{f})\})}=0.
\]
\noindent We say that $T\sim\hat{T}$ if and only if $T\sim_{\vec{f}}\hat{T}$
for all $\vec{f}.$
\end{definition}

That is, two tests are equivalent if and only if, given an ordered
 pair of forecasting strategies, there is zero probability according
to each forecaster that the tests will converge to different propensities.
\begin{claim}
\label{Claim equivalence relation-1}The relation $\sim$ is an equivalence
relation on $\mathrm{\top}:=\{T:\ T-cardinal\ comparison\ test\}.$
\end{claim}

The proof of Claim \ref{Claim equivalence relation-1} is relegated
to Appendix \ref{sec:Appendix Missing-proofs-1}. The next theorem
asserts that, up to an equivalence class representative, there exists
a unique anonymous reasonable and error-free cardinal comparison test. That is, any anonymous
test $T\nsim T_{{\cal D}}$ which is reasonable, admits an error.
To this end, we will show that any $T\nsim T_{{\cal D}}$ can be associated
with a pair of forecasting strategies for which the error-free condition
fails. More importantly, the power of the theorem stems from the premise
that $T$ admits an error at any pair $\vec{f}$ whenever $T\nsim_{\vec{f}}{\cal D}.$

Before proceeding, we make the observation that Definition \ref{Def: equivalence tests}
can be stated equivalently by the next lemma which is invoked in our
adjacent uniqueness theorem proof.

\vspace{0cm}

\begin{lemma}
\textup{\label{Lemma: equivalence definition to equivalence}Let $\vec{f}:=(f,g)\in F\times F.$
Then {\em$T\sim_{\vec{f}}\hat{T}$ } if and only if for all $\epsilon\in(0,1)\cap\mathbb{Q}$}
\textup{
\[
f((L_{T,\epsilon}^{\vec{f}}\cap R_{\hat{T},\epsilon}^{\vec{f}})\cup(L_{\hat{T},\epsilon}^{\vec{f}}\cap R_{T,\epsilon}^{\vec{f}}))=g((L_{T,\epsilon}^{\vec{f}}\cap R_{\hat{T},\epsilon}^{\vec{f}})\cup(L_{\hat{T},\epsilon}^{\vec{f}}\cap R_{T,\epsilon}^{\vec{f}}))=0.
\]
}
\end{lemma}
\vspace{0cm}
The proof of Lemma \ref{Lemma: equivalence definition to equivalence}
is supplemented to Appendix \ref{sec:Appendix Missing-proofs-1}.
\begin{theorem}
\textup{\label{Th: first main theorem uniqueness}Let $T$ be an anonymous
and reasonable cardinal comparison test. If $T\nsim{\cal D}$ then $T$ is not  error-free.}
\end{theorem}
\proof{Proof.}\
Assume by contradiction that $T$ is error-free. Let $\vec{f}:=(f,g)$
be such that $T\nsim_{\vec{f}}{\cal D}$; then from Lemma \ref{Lemma: equivalence definition to equivalence}
there exits $\epsilon\in(0,1)$ such that (w.l.o.g. for $f$)
\[
f((L_{{\cal D},\epsilon}^{\vec{f}}\cap R_{T,\epsilon}^{\vec{f}})\cup(L_{T,\epsilon}^{\vec{f}}\cap R_{{\cal D},\epsilon}^{\vec{f}}))>0.
\]
\noindent We shall consider the following cases which result in a
contradiction.

Case 1: $f(L_{{\cal D},\epsilon}^{\vec{f}}\cap R_{T,\epsilon}^{\vec{f}})>0.$
Assume that $\epsilon\in(\frac{1}{2},1)$ and observe that since $L_{{\cal D},\epsilon}^{\vec{f}}=\stackrel[n\in\mathbb{N}:n>\lceil\frac{1}{1-\epsilon}\rceil]{\infty}{\bigcup}L_{{\cal D},\epsilon+\frac{1}{n}}^{\vec{f}}$
and $f$ is monotone with respect to inclusion, there exist $\epsilon<\epsilon_{1}$
such that $f(\hat{A_{1}}:= L_{{\cal D},\epsilon_{1}}^{\vec{f}}\cap R_{T,\epsilon}^{\vec{f}})>0.$

\noindent By Proposition \ref{Prop 2: D is  error-free}, ${\cal D}$
is error-free where $\hat{A_{1}}\subset L_{{\cal D},\epsilon_{1}}^{\vec{f}}$;
hence
\[
g(\hat{A_{1}})\leq(\frac{1-\epsilon_{1}}{\epsilon_{1}})f(\hat{A_{1}})<(\frac{1-\epsilon}{\epsilon})f(\hat{A_{1}}).
\]
\noindent In addition, by the assumption $T$ is reasonable hence
\[
f(\hat{A_{1}}\cap L_{T,\epsilon}^{\vec{f}})>0,
\]
\noindent which yields a contradiction since $R_{T,\epsilon}^{\vec{f}},L_{T,\epsilon}^{\vec{f}}$
are disjoint sets.

For $\epsilon\in(0,\frac{1}{2})$ note that $R_{T,\epsilon}^{\vec{f}}=\stackrel[n\in\mathbb{N}:n>\lceil\frac{1}{\epsilon}\rceil]{\infty}{\bigcup}R_{T,\epsilon-\frac{1}{n}}^{\vec{f}},$
and hence there exists $\epsilon_{2}<\epsilon$ such that $f(A_{2}:=\{L_{{\cal D},\epsilon}^{\vec{f}}\cap R_{T,\epsilon_{2}}^{\vec{f}}\})>0$.
By the assumption $T$ is an error-free test, hence
\[
f(\hat{A_{2}})\leq(\frac{\epsilon_{2}}{1-\epsilon_{2}})g(\hat{A_{1}})<(\frac{\epsilon}{1-\epsilon})g(\hat{A_{2}}).
\]
\noindent In addition, by Proposition \ref{Prop 2: D is  error-free},
${\cal D}$ is reasonable hence 
\[
g(\hat{A_{2}}\cap R_{{\cal D},\epsilon}^{\vec{f}})>0,
\]
\noindent which yields a contradiction since $R_{{\cal D},\epsilon}^{\vec{f}},L_{{\cal D},\epsilon}^{\vec{f}}$
are disjoint sets.

Case 2: $f(\hat{A_{3}}:= L_{T,\epsilon}^{\vec{f}}\cap R_{{\cal D},\epsilon}^{\vec{f}})>0.$
Assume (w.l.o.g) that $\epsilon\in(\frac{1}{2},1).$ By the assumption $T$
is an error-free test where, by Proposition \ref{Prop 2: D is  error-free},
${\cal D}$ is reasonable, therefore, the contradiction 
\[
g(\hat{A_{3}}\cap L_{T,\epsilon}^{\vec{f}})>0,
\]
follows analogously from Case 1 and hence omitted.\Halmos
\endproof

\section{{\label{sec:Independent-axioms}Independence of axioms}}\

The notions of error-freeness and reasonableness which were introduced
in Subsection \ref{subsec:Desirable-Properties} are not related;
obviously, as the sets: $\{T=\frac{1}{2}\},\ R_{T,\epsilon}^{\vec{f}},\ L_{T,1-\epsilon}^{\vec{f}}$
are disjoint, inequality \eqref{eq: error free condition} is satisfied
trivially and hence the constant fair test, $T_{t}(\omega,\vec{f})\equiv1/2,$
is error-free and is not reasonable. Using the result of Theorem \ref{Thorem:  D is error-free and reasonable},
the next example illustrates that reasonableness does not imply error-freeness.
\begin{example}
Let $\overrightarrow{h_{2}},\overleftarrow{h_{2}}$ be play paths
composed of the realization $\tilde{\omega}:=(1,1,1,,,),$ and pairs
of forecasts along $\tilde{\omega}$ which, from day two onward, are
shown to have similar forecasts according to an iid distribution with
parameter $1,$ where on day one, one forecast assigns 1 to the outcome
1 whereas the other assigns half. Let $\overrightarrow{h_{2}^{t}},\overleftarrow{h_{2}^{t}}$
be the corresponding uniquely induced $t$ - history and consider
the following test:
\[
T_{t}(\omega,\vec{f})=\begin{cases}
\begin{array}{l}
{\cal D}_{t}(\omega,\vec{f}),\\
0,\\
1,
\end{array} & \begin{array}{l}
other\\
(\omega,\vec{f})^{t}=\overrightarrow{h_{2}^{t}}\\
(\omega,\vec{f})^{t}=\overleftarrow{h_{2}^{t}}.
\end{array}\end{cases}
\]
\end{example}
\begin{claim}
$T$ is anonymous and reasonable but is not error-free.
\end{claim}
\proof{Proof.}

Since ${\cal D}$ is anonymous and $T_{t}(\omega,f,g)=1-T_{t}(\omega,g,f)$ whenever $(\omega,\vec{f})^{t}$ equals $\overrightarrow{h_{2}^{t}}$ or $\overleftarrow{h_{2}^{t}}$, it follows that $T$ is anonymous. Further, let $\vec{\tilde{f}}=(\tilde{f},\tilde{g})$ be such that $(\tilde{\omega},\vec{\tilde{f}})=\overrightarrow{h_{2}}$. Since $\{\tilde{\omega}\}=R_{T,\epsilon}^{\vec{\tilde{f}}}$ for all $\epsilon\in(0,\frac{1}{2})$ one has
\[\tilde{f}(\{\tilde{\omega}\}\cap R_{T,\frac{1}{3}}^{\vec{\tilde{f}}})=1>\frac{1}{2}\tilde{g}(\{\tilde{\omega}\}\cap R_{T,\frac{1}{3}}^{\vec{\tilde{f}}})
\]
and hence $T$ is not error-free. To verify that $T$ is a reasonable test let $\vec{f}:=(f,g)$, a measurable set $A$, and $\epsilon\in(0,\frac{1}{2})$. If $(\tilde{\omega},\vec{f})\neq\overrightarrow{h_{2}}$ and $(\tilde{\omega},\vec{f})\neq\overleftarrow{h_{2}}$, then by the construction $T_{t}(\cdot,\vec{f})\equiv{\cal D}_{t}(\cdot,\vec{f})$ and since by Proposition \ref{Prop 3 : D is  reasonable} ${\cal D}$ is a reasonable test, condition \eqref{eq:reasonableness condition} is satisfied. Now, observe that if $(\tilde{\omega},\vec{f})=\overrightarrow{h_{2}}$ and $\tilde{\omega}\in A$ then $f(A)=1>\frac{\epsilon}{1-\epsilon}g(A)$ which rules out the left hand-side of condition \eqref{eq:reasonableness condition}. If, on the other hand, $(\tilde{\omega},\vec{f})=\overrightarrow{h_{2}}$ and $\tilde{\omega}\notin A$, then $g(A)>0$ implies that $\hat{\omega}:=(0,1,1,,,)\in A$ where $g(A\setminus\hat{\omega})=0$, in which case $f(A)=0$ and $T_{t}(\hat{\omega},\vec{f})\equiv{\cal D}_{t}(\hat{\omega},\vec{f})=0$. Hence, since by Proposition \ref{Prop 3 : D is  reasonable}  ${\cal D}$ is reasonable we have $g(\hat{\omega}\cap R_{T,\epsilon}^{\vec{f}})=g(A\cap R_{T,\epsilon}^{\vec{f}})=g(A\cap R_{{\cal D},\epsilon}^{\vec{f}})>0$. For the remaining case, note that if $(\tilde{\omega},\vec{f})=\overleftarrow{h_{2}}$ then either $\tilde{\omega}\in A$, in which case $f(A)\geq\frac{1}{2}>\frac{\epsilon}{1-\epsilon}g(A)$, or $\tilde{\omega}\notin A$ yielding that $g(A)=0$. Since the case for which $\epsilon\in(\frac{1}{2},1)$ is proven analogously, the result follows.\Halmos
\endproof

\citet{Al-Najjar-2008} introduce an alternative cardinal comparison
test:
\[
L_{t}(\omega,f,g)=\begin{cases}
\begin{array}{l}
0,\\
0.5,\\
1,
\end{array} & \begin{array}{l}
\frac{g(\omega^{t})}{f(\omega^{t})}>1\\
other\\
\frac{g(\omega^{t})}{f(\omega^{t})}<1.
\end{array}\end{cases}
\]
Note that this test differs from ${\cal D}$ whenever the likelihood
ratio is high but finite. In our case, the test does not prefer any
expert but provides a relative ranking, whereas the likelihood ratio test, $L$, does.\footnote{A different existing test, which was introduced in \citet{Feinberg-Stewart-2008},
is the cross-calibration test which is discussed in the introduction.
However, it turns out that this test does not naturally induce a cardinal
comparison test; rather than ranking the experts, this test outputs
a binary verdict (pass/fail) for each of the two experts separately
and hence may rule out anonymity. Moreover, it can be shown that any
cardinal comparison test which naturally ranks an expert according
to its empirical frequency would fail to be reasonable if an expert,
who had calibrated only along one profile and had failed along all
others, is tested against an informed expert .} 
\begin{claim}
$L$ is anonymous and reasonable and is not error-free.
\end{claim}
\proof{Proof.}
Let $\epsilon\in(\frac{1}{2},1)$. Let $g$ be a forecasting strategy which deterministically predicts $\tilde{\omega}$, and let $f$ be such that it predicts $(1-\epsilon)$ at day one and meets $g$ from day two onward regardless of any past history. Note that whenever $f$ is assumed to be the true measure, then $L_{t}(\tilde{\omega},f,g)=\frac{1}{1-\epsilon}>1$ for all $t>0$ and so expert $g$ is determinstically ranked by 1 along $(\tilde{\omega},\vec{f})$ yielding $\tilde{\omega}\in L_{L,\epsilon}^{\overrightarrow{f}}$. A simple calculation shows that
\[g(\tilde{\omega}\cap L_{L,\epsilon}^{\overrightarrow{f}})>\frac{1-\epsilon}{\epsilon}f(\tilde{\omega}\cap L_{L,\epsilon}^{\overrightarrow{f}})
\]
as $g(\tilde{\omega})=1$. Since $\epsilon$ is taken arbitrarily, the following important conclusion can be drawn: not only is $\text{\ensuremath{L}}$ not error-free but it admits an arbitrarily large error. The fact that L is reasonable follows directly from Proposition \ref{Prop 3 : D is  reasonable}.\Halmos
\endproof

\section{{Decisiveness in finite time}}\

In this section we provide a natural sufficient condition for which
a tester achieves a higher level of confidence in favor of the informed
forecaster with any desired degree of precision in some fixed finite
time. To this end, we show the existence of a uniform bound on the
rate at which a cardinal comparison test converges. Consider expert $f's$ point of view.
Not only should he maintain that, whenever expert $g$'s forecasts
are different from his, then he should eventually be ranked higher
than him, but if expert $g$'s forecasts are relatively far, then
this should essentially happen uniformly fast. Indeed, as we show
in this section, this holds for our finite derivative test. This observation
tightly builds on a theory of active supermartingales due to \citet{Fudenberg-Levine-1992}.

\vspace{0cm}

To determine whether a test is `almost' certain about a forecaster
requires the two forecasters to provide significantly different forecasts
as captured by the following definition:

\begin{definition}
A pair of forecasting strategies $\vec{f}:=(f,g)$ is {\em $\epsilon-close$
} along $\omega$ at period $t>0,$ if 
\[
|f((\omega,\vec{f})^{t-1})[\omega_{t}]-g((\omega,\vec{f})^{t-1})[\omega_{t}]|<\epsilon
\]
\end{definition}

The next theorem asserts that, given an arbitrarily small $\epsilon>0,$
there exists a finite uniform bound, $K,$ which is independent of
any pair of forecasting strategies, such that if the forecasts of
the uninformed expert are sufficiently different from those of the
informed one in more than $K$ periods, then the finite derivative
test, ${\cal D}$, will eventually settle on the informed expert with
a high level of confidence. In the latter scenario, it furthermore
surprisingly asserts that, given any sufficiently large time $n$,
${\cal D}_{n}$ ranks the informed expert higher than $(1-\epsilon)$
and up to $\epsilon$ - amount of accuracy as it would have ranked
had it continued to rank the expert following his test to infinity.
\begin{theorem}
\textup{\label{Th: second main theorem (F=000026L)}For all $0<\epsilon<1$
there exists $K=K(\epsilon)$ such that for all $\vec{f}:=(f,g),$
and for all $n>0$, there is a set of which the probability according
to $f$ is at least $(1-\epsilon)$ such that for any $\omega$ in
that set: }
\begin{enumerate}
\item \textup{\label{enu:Either--is condition 1-1}Either $\vec{f}$ is
$\epsilon-close$ along $\omega$ in all but $K$ periods in $\{1...n\}$
or}
\item \textup{\label{enu:-is-condition 2-1}$\omega\in L_{{\cal D},1-\epsilon}^{\vec{f}}$.
Furthermore, $|{\cal D}_{t}(\omega,\vec{f})-{\cal D}_{n}(\omega,\vec{f})|<\epsilon$
for all $t\geq n$. }
\end{enumerate}
\end{theorem}
\vspace{0cm}

In words, with high probability, given any sufficiently large $n$
and any sufficiently small $\epsilon$, the only reason that the tester
is not `almost' settled on the correct forecaster at time $n$ is
because the uninformed expert made excellent predictions along the
play path. Moreover, Theorem \ref{Th: second main theorem (F=000026L)}
is universal in the following manner: The bound on the number of periods
in which the two experts\textquoteright{} forecasts must be different,
$K$, for the finite derivative test to rank the informed one higher,
depends on the required level of accuracy, but is independent of any
pair of forecasting strategies, $f$ or $g$.

The proof of Theorem \ref{Th: second main theorem (F=000026L)} is
relegated to Appendix \ref{sec:Appendix Missing-proofs-1}. Nevertheless
let us briefly provide some technical intuition. At the heart of the
proof of Theorem \ref{Th: second main theorem (F=000026L)} lies a
theorem due to regarding the rate of decrease of {\em active} supermartingales.
Consider an abstract setting with a probability measure $P$ in $\Delta(\Omega^{\infty})$
and a filtration $\{{\cal G}_{t}\}_{t=1}^{\infty}$. 
\begin{definition}
A $({\cal G}_{t})$ - adapted, real-valued process $\tilde{{\cal D}}:=\{\tilde{{\cal D}}_{t}\}_{t=0}^{\infty}$
is called a {\em supermartingale} under $P$ if

\end{definition}
\begin{enumerate}

\item $E|\tilde{{\cal D}}_{t}|<\infty$ for all $t>0$;
\item $E[\tilde{{\cal D}}_{t}|{\cal G}_{s}]\leq\tilde{{\cal D}}_{t}$ for
all $s\leq t$, $P-a.s.$
\end{enumerate}

Intuitively, a supermartingale is a process that decreases on average.
The proof of Theorem \ref{Th: second main theorem (F=000026L)} implies
that the finite  derivative test is associated with a supermartingale
property with respect to the natural filtration which is defined in
Section \ref{sec:Model}. Let us further consider the following class
of supermartingales called active supermartingales. This notion was
first introduced in \citet{Fudenberg-Levine-1992} who studied reputations
in infinitely repeated games:
\begin{definition}
A non-negative supermartingale $\tilde{{\cal D}}$ is {\em active}
with activity $\psi\in(0,1)$ under $P$ if 
\[
P(\{\omega:\ |\frac{\tilde{{\cal D}}_{t}(\omega)}{\tilde{{\cal D}}_{t-1}(\omega)}-1|>\psi\}|\tilde{\omega}^{k-1})>\psi
\]
\noindent for $P$ - almost all histories $\tilde{\omega}^{t-1}$
such that $\tilde{{\cal D}}_{t-1}(\tilde{\omega})>0$.
\end{definition}
In other words, a supermartingale  has activity $\psi$ if the probability
of a jump of size $\psi$ at time $t$ exceeds $\psi$ for almost
all histories. Note that $\tilde{{\cal D}}$ being a supermartingale,
is weakly decreasing in expectations. Showing that it is active implies
that $\tilde{{\cal D}}_{t}$ substantially goes up or down relative
to $\tilde{{\cal D}}_{t-1}$ with probability bounded away from zero
in each period. \citet{Fudenberg-Levine-1992}, Theorem A.1, showed
the following remarkable result
\begin{theorem}[\citet{Fudenberg-Levine-1992}]
\textup{\label{Th A.1 (F=000026L)}For every $\epsilon>0,\ \psi\in(0,1),$
and $0<\underset{-}{D}<1$ there is a time $K<\infty$ such that }
\textup{
\[
P(\{\omega:\ \underset{t>K}{sup}\tilde{{\cal D}}_{t}(\omega)\leq\underset{-}{D}\})\geq1-\epsilon
\]
}
\noindent \textup{for every active supermartingale $\{\tilde{{\cal D}}_{t}\}$
with $\tilde{{\cal D}}_{0}\equiv1$ and activity $\psi.$ }
\end{theorem}
Theorem \ref{Th A.1 (F=000026L)} asserts that if $\tilde{{\cal D}}$
is an active supermartingale with activity $\psi$, then there is
a fixed time $K$ by which, with high probability, $\tilde{{\cal D}}_{t}$
drops below $\underset{-}{D}$ and remains below $\underset{-}{D}$
for all future periods. It should be noted that the power of the theorem
stems from the fact that the bound, $K$, depends solely on the parameters
$\epsilon>0,\ \psi$ and $\underset{-}{D}$ , and is otherwise independent
of the underlying stochastic process $P$. 

\vspace{0cm}

We exploit the active supermartingale property in a different way.
In the context of cardinal comparison testing, we consider two strategies,
one for each expert, which are updated using Bayes rule. Given sufficiently
small $\epsilon>0,$ our comparative test ranks an expert depending
on whether the posterior odds ratio is above or below $\epsilon$.
The active supermartingale result implies that there is a uniform
bound (independent of neither the length of the game nor the true
distribution) on the number of periods where the uninformed expert
can be substantially wrong, without being detected, such that if this
bound is exceeded, the probability that the tester ranks high the
uninformed expert is small. 

\vspace{0cm}

\section{{Concluding remarks}}\

The paper proposes a normative approach to the challenge of comparing
between two forecasters who repeatedly provide probabilistic forecasts.
The paper postulates three basic norms: anonymity, error-freeness
and reasonableness and provides a cardinal comparison test, the finite
derivative test, that complies with them. It also shows that this
test is essentially unique. Finally, it shows that the test converges
fast and hence is meaningful in finite time. In the future we hope
to extend our results to settings with more than two forecasters and
study alternative sets of norms.

\subsection{{Implications}}\

The approach taken in this paper can be considered as a contribution
to the hypothesis testing literature in statistics where a forecaster
is associated with a hypothesis. In this context we propose a hypothesis
test that complies with a set of fundamental properties which we refer
to as axioms. In contrast, a central thrust for the hypothesis testing
literature (for two hypotheses) is the pair of notions of significance
level and power of a test. In that literature one hypothesis is considered
as the null hypothesis while the other serves as an alternative. A
test is designed to either reject the null hypothesis, in which case
it accepts the alternative, or fail to reject it (a binary outcome).
The significance level of a test is the probability of rejecting the
null hypothesis whenever it is correct (type-1 error) while the power
of the test is the probability of rejecting the null hypothesis assuming
the alternative one is correct (the complement of a type-2 error). 

In contrast with the aforementioned binary outcome that is prevalent
in the hypothesis testing literature we allow, in addition, for an
inconclusive outcome. Recall the celebrated Neyman-Pearson lemma which
characterizes a test with the maximal power subject to an upper bound
on the significance level. The possibility of an inconclusive (ranking)
outcome, in our framework, allows us to design a test where both type-1
and type-2 errors have relatively low probability.\footnote{Note that we abuse the statistical terminology. In statistics the
notion of rejection is always used in the context of the null hypothesis.
In our model, we assume symmetry between the alternatives and so we
discuss rejection also in the context of the alternative hypothesis.
As a consequence, an error of type-1 is defined as the probability
of accepting the alternative hypothesis whenever the null hypothesis
is correct, and symmetrically, an error of type-2 is the probability
of accepting the null hypothesis whenever the alternative one is correct.}

Interestingly, the test proposed in the Neyman-Pearson lemma, similar
to ours, also hinges on the likelihood ratio.\footnote{The test proposed in the Neyman-Pearson lemma rejects the null hypothesis
whenever the likelihood ratio falls below some positive threshold.} In our approach we, a priori, treat both hypotheses symmetrically.
In the statistics literature, however, this is not the case and the
null hypothesis is, in some sense, the status quo hypothesis. This
asymmetry is manifested, for example, in the Neyman-Pearson lemma.

Note that in order to design a test that complies with a given significance
level and a given power one must know the full specification of the
two hypotheses. This is in contrast with our test which is universal,
in the sense that it does not rely on the specifications of the two
forecasts. Finally, let us comment that whereas hypothesis testing
is primarily discussed in the context of a finite sample, typically
from some iid distribution, our framework allows for sequences of
forecasts that are dependent on past outcomes as well as past forecasts
of the other expert.

\vspace{0cm}

% Acknowledgments here
\section*{Acknowledgments}\

Smorodinsky gratefully acknowledges the United States-Israel Binational Science Foundation and the National Science Foundation (grant 2016734),
the German-Israel Foundation (grant I-1419-118.4/2017), 
the Ministry of Science and Technology (grant 19400214), 
the Technion VPR grants, and the Bernard M. Gordon Center for Systems Engineering at the Technion.

%\section*{{\normalsize{} \center{APPENDIX}
%}}

%\appendix

\begin{APPENDICES}

\section{\label{Appendinx sec:On-ideal-tests}On ideal tests}\
%\section{Appendinx}
%\subsection{\label{Appendinx sec:On-ideal-tests}On ideal tests}\

Recall that an error-free test eliminates the necessity of pointing
out the less informed expert. A stronger and more appealing property
is to point out the better informed expert, in which case the tester
eventually settles on one forecaster as being better than the other.
We consider tests that exhibit such a property as ideal. Formally,
\begin{definition}
\label{Def: T is decisive-1-1-2}$T$ is {\em decisive} on $f$ at
$(\omega,\vec{f})$ (respectively, g) if $T_{t}(\omega,\vec{f})\longrightarrow1$
(respectively, $(1-T_{t}(\omega,\vec{f}))\longrightarrow1$).
\end{definition}
\vspace{0cm}

For a given $T,\vec{f},$ we denote by
\[
A_{T,f}^{\vec{f}}:=\{\omega\colon\ T\ is\ decisive\ on\ f\ at\ (\omega,\vec{f})\},
\]
\noindent to be the measurable set of realizations (in $L_{T,N}^{\vec{f}}$)
for which $T$ is decisive on $f$ at $(\omega,\vec{f}).$
\begin{definition}
A test $T$ is {\em ideal with respect to $A\subseteq F$} if for
all $\vec{f}:=(f_{,}g\neq f)\in A\times A$
\[
f(A_{T,f}^{\vec{f}})=g(A_{T,g}^{\vec{f}})=1.
\]
\noindent It is called {\em ideal} if it is ideal with respect
to $F.$ 
\end{definition}
In other words, whenever the left expert knows the actual data-generating
process and the right expert does not, an ideal test will surely identify
the informed expert.

Trivially, any ideal test with respect to a subset of forecasts $A$
is also error-free with respect to the same set. The following is
a straightforward corollary of Theorem \ref{Thorem:  D is error-free and reasonable}. 
\begin{corollary}
\textup{There exists no ideal test with respect to a set of forecasts
$A$ whenever it contains two forecasts which induce measures, one
of which is absolutely continuous with respect to the other.}
\end{corollary}
This immediately entails: 
\begin{corollary}
\textup{There exists no ideal test. }
\end{corollary}
However, whenever $A$ contains no such pair of forecasts, then an
ideal test does exist. To prove this we must first accurately define
the notion of mutually singular forecasts. 
\begin{definition}
Two forecasting strategies, $\vec{f}=(f,g\neq f)\in F,$ are said
to be mutually singular with respect to each other, if there exist
two disjoint sets 
\[
C_{f}^{\vec{f}},C_{g}^{\vec{f}}\subset(\Omega\times\Delta(\Omega)\times\Delta(\Omega))^{\infty}
\]
such that\footnote{Recall that $\vec{f}$ induces a unique play path $(\omega,\vec{f})$$.$}
\[
f(\{\omega:\ (\omega,\vec{f})\in C_{f}^{\vec{f}}\})=g(\{\omega:\ (\omega,\vec{f})\in C_{g}^{\vec{f}}\})=1.
\]
A set $A\subseteq F$ is pairwise mutually singular if for all $\vec{f}=(f,g)\neq f)\in A$,
$f,g$ are mutually singular with respect to each other. 
\end{definition}
The next lemma asserts that a reasonable test is able to perfectly
distinguish between far measures which are induced from forecasting
strategies which are mutually singular with respect to each other. 
\begin{lemma}
\textup{\label{lem:reasonable implies m.s}Let $f,g\neq f\in F$ which
are mutually singular with respect to each other. If $T$ is reasonable
then}
\textup{
\[
f(A_{T,f}^{\vec{f}})=g(A_{T,g}^{\vec{f}})=1.
\]
}
\end{lemma}
\noindent The proof of Lemma \ref{lem:reasonable implies m.s} is
relegated to Appendix \ref{sec:Appendix Missing-proofs-1}. It should
be noted that Lemma \ref{lem:reasonable implies m.s} holds even for
$T$ which is not error-free. 

\vspace{0cm}

\noindent The next theorem provides a necessary and sufficient condition
for the existence of an ideal test over sets. 
\begin{theorem}
\textup{There exists an anonymous ideal test with respect to $A$ if and only if
A is pairwise mutually singular.}
\end{theorem}
\proof{Proof.}\

$\Longleftarrow$ Directly follows from Lemma \ref{lem:reasonable implies m.s}
and Proposition \ref{Prop 3 : D is  reasonable}.

$\Longrightarrow$ Let $T$ be an ideal anonymous test with respect
to a set $A$. Let $\vec{f}:=(f,g\neq f)\in A\times A$ and denote
\[
C_{f}^{T}:=\{(\omega,\vec{f}):\omega\in A_{T,f}^{\vec{f}}\},\ C_{g}^{T}:=\{(\omega,\vec{f}):\omega\in A_{T,g}^{\vec{f}}\}.
\]
\noindent Since $A_{T,f}^{\vec{f}},\ A_{T,g}^{\vec{f}}$ are disjoint,
it follows that $C_{f}^{T},\ C_{g}^{T}$ are disjoint where $T$ ideal
yields 
\[
f(\{\omega:(\omega,\vec{f})\in C_{f}^{T}\})=f(A_{T,f}^{\vec{f}})=g(A_{T,g}^{\vec{f}})=g(\{\omega:(\omega,\vec{f})\in C_{g}^{\vec{f}}\})=1.
\]        
\Halmos
\endproof
We conclude the paper with an example of an ideal test over a domain
of mutually singular forecasts:
\begin{example}Let 
\[
A_{iid}\times A_{iid}:=\{\vec{f}:=(f,g):\text{there exist }a_{f},a_{g}\in[0,1]\text{ s.t }f(\omega^{t})[1]\equiv a_{f},\ g(\omega^{t})[1]\equiv a_{g}\text{ for all }\omega\in\Omega^{\infty}\}
\] and for $\omega\in\Omega^{\infty}$ denote the average realization
by
\[
a_{\omega}:=\underset{t\rightarrow\infty}{lim}\left(\frac{\stackrel[n=1]{t}{\sum}1_{\{\omega_{n}=1\}}}{t}\right)
\]
(whenever the limit exists). Let $\vec{f}\in A_{iid}\times A_{iid}$ such that $a_{f}\neq a_{g}$ and observe that for any $\omega$ 
\[
a_{\omega}=a_{f}\iff\underset{t\rightarrow\infty}{lim}D_{f}^{t}g(\omega)=0\iff\underset{t\rightarrow\infty}{lim}{\cal D}_{t}(\omega,\vec{f}))=1.
\]
\noindent Since the induced measures $f,g$ are iid with different
parameters, a mere application of the law of large numbers yields 
\[
f(A_{{\cal D},f}^{\vec{f}})=1\ and\ g(A_{{\cal D},f}^{\vec{f}})=0,
\]
showing that ${\cal D}$ is ideal with respect to $A_{iid}$. 
\end{example}

\setcounter{equation}{0} \renewcommand{\theequation}{B.\arabic{equation}}
\section{{\label{sec:Appendix Missing-proofs-1}Missing proofs}}

\proof{\textbf{Proof of Lemma \ref{Lemma: equivalence definition to equivalence}}.}
Observe that for all $\omega\in\{\text{T}\neq\hat{T}\}$ there exists
$\epsilon\in(0,1)\cap\mathbb{Q}$ such that either $\hat{T}(\omega,\vec{f})<\epsilon<T(\omega,\vec{f})$
or $T(\omega,\vec{f})<\epsilon<\hat{T}(\omega,\vec{f}).$ We thus
have

\begin{equation}
\{\hat{T}<T\}=\underset{\epsilon\in(0,1)\cap\mathbb{Q}}{\bigcup}\{\hat{T}<\epsilon<T\}=\underset{\epsilon\in(0,1)\cap\mathbb{Q}}{\bigcup}(L_{T,\epsilon}^{\vec{f}}\cap R_{\hat{T},\epsilon}^{\vec{f}})\label{eq: T^< T}
\end{equation}

\noindent as well as

\begin{equation}
\{\hat{T}>T\}=\underset{\epsilon\in(0,1)\cap\mathbb{Q}}{\bigcup}\{T<\epsilon<\hat{T}\}=\underset{\epsilon\in(0,1)\cap\mathbb{Q}}{\bigcup}(L_{\hat{T},\epsilon}^{\vec{f}}\cap R_{T,\epsilon}^{\vec{f}}).\label{eq: T^> T}
\end{equation}

$\Longleftarrow$Assume by contradiction that $T\nsim_{\vec{f}}\hat{T}$
and observe that since $\{\text{T}\neq\hat{T}\}=\{\hat{T}<T\}\cup\{\hat{T}>T\}$
it follows that (w.l.o.g. for $f)$ either $f(\{\hat{T}<T\})>0$ or
$f(\{\hat{T}>T\})$. If $f(\{\hat{T}<T\})>0$ then the most right
equality of \eqref{eq: T^< T} implies that there exists $\epsilon'\in(0,1)\cap\mathbb{Q}$
such that $f(L_{T,\epsilon'}^{\vec{f}}\cap R_{\hat{T},\epsilon'}^{\vec{f}})>$0
yielding a contradiction (similarly whenever $f(\{\hat{T}<T\})>0$).

$\Longrightarrow$ Assume by contradiction that (w.l.o.g. for $f)$
$f((L_{T,\epsilon'}^{\vec{f}}\cap R_{\hat{T},\epsilon'}^{\vec{f}})\cup(L_{\hat{T},\epsilon'}^{\vec{f}}\cap R_{T,\epsilon'}^{\vec{f}}))>0$
for some $\epsilon'\in(0,1)\cap\mathbb{Q}.$ Therefore, either $f(\underset{\epsilon\in(0,1)\cap\mathbb{Q}}{\bigcup}(L_{T,\epsilon}^{\vec{f}}\cap R_{\hat{T},\epsilon}^{\vec{f}}))>0$
or $f(\underset{\epsilon\in(0,1)\cap\mathbb{Q}}{\bigcup}(L_{\hat{T},\epsilon}^{\vec{f}}\cap R_{T,\epsilon}^{\vec{f}}))>0.$
In which case, using \eqref{eq: T^< T} and \eqref{eq: T^> T}, we
conclude $f(\{\hat{T}<T\}\cup\{\hat{T}>T\})>0$ which contradicts
that $\text{T}\sim_{\vec{f}}\hat{T}.$\Halmos
\endproof

\proof{\textbf{Proof of Claim \ref{Claim equivalence relation-1}}.}
Let $T,T_{1},T_{2}\in\top,\ \vec{f}\in F\times F$.

\noindent Reflexivity: Applying Lemma \ref{Lemma: equivalence definition to equivalence}
it is readily seen that $f(L_{T,\epsilon}^{\vec{f}}\cap R_{T,\epsilon}^{\vec{f}})=g(L_{T,\epsilon}^{\vec{f}}\cap R_{T,\epsilon}^{\vec{f}})=0$
as $L_{T,\epsilon}^{\vec{f}},R_{T,\epsilon}^{\vec{f}}$ are disjoint
sets for all $\epsilon\in(0,1)$. Thus, $T\sim_{\vec{f}}T.$

\noindent Anonymity: From Lemma \ref{Lemma: equivalence definition to equivalence}
we obtain that for all $\epsilon\in(0,1)\cap\mathbb{Q},$
\[
\begin{array}{ll}
T_{1}\sim_{\vec{f}}T_{2} & \iff f((L_{T_{1},\epsilon}^{\vec{f}}\cap R_{T_{2},\epsilon}^{\vec{f}})\cup(L_{T_{2},\epsilon}^{\vec{f}}\cap R_{T_{1},\epsilon}^{\vec{f}}))=g((L_{T_{1},\epsilon}^{\vec{f}}\cap R_{T_{2},\epsilon}^{\vec{f}})\cup(L_{T_{2},\epsilon}^{\vec{f}}\cap R_{T_{1},\epsilon}^{\vec{f}}))\\
 & =f((L_{T_{2},\epsilon}^{\vec{f}}\cap R_{T_{1},\epsilon}^{\vec{f}})\cup(L_{T_{1},\epsilon}^{\vec{f}}\cap R_{T_{2},\epsilon}^{\vec{f}}))=g((L_{T_{2},\epsilon}^{\vec{f}}\cap R_{T_{1},\epsilon}^{\vec{f}})\cup(L_{T_{1},\epsilon}^{\vec{f}}\cap R_{T_{2},\epsilon}^{\vec{f}}))\\
 & \iff T_{2}\sim_{\vec{f}}T_{1}.
\end{array}
\]
\noindent Transitivity: Suppose by contradiction that $T_{1}\sim_{\vec{f}}T,$
and $T\sim_{\vec{f}}T_{2}$ where $T_{1}\nsim_{\vec{f}}T_{2}.$ Then
from Lemma \ref{Lemma: equivalence definition to equivalence} (wl.o.g.
for $f)$ we are provided with $\bar{\epsilon}\in(0,1)$ such that
\[
f((L_{T_{1},\bar{\epsilon}}^{\vec{f}}\cap R_{T_{2},\bar{\epsilon}}^{\vec{f}})\cup(L_{T_{2},\bar{\epsilon}}^{\vec{f}}\cap R_{T_{1},\bar{\epsilon}}^{\vec{f}}))>0,
\]
\noindent where for all $\epsilon\in(0,1),$

\begin{equation}
\begin{array}{l}
T_{1}\sim_{\vec{f}}T\Longrightarrow f((L_{T_{1},\epsilon}^{\vec{f}}\cap R_{T,\epsilon}^{\vec{f}})\cup(L_{T,\epsilon}^{\vec{f}}\cap R_{T_{1},\epsilon}^{\vec{f}}))=0,\\
T\sim_{\vec{f}}T_{2}\Longrightarrow f((L_{T,\epsilon}^{\vec{f}}\cap R_{T_{2},\epsilon}^{\vec{f}})\cup(L_{T_{2},\epsilon}^{\vec{f}}\cap R_{T,\epsilon}^{\vec{f}}))=0.
\end{array}\label{eq: Transitivity condition}
\end{equation}

Case 1: $f(L_{T_{1},\bar{\epsilon}}^{\vec{f}}\cap R_{T_{2},\bar{\epsilon}}^{\vec{f}})>0.$
Note that,
\noindent 
\[
\begin{array}{cl}
f(L_{T_{1},\bar{\epsilon}}^{\vec{f}}\cap R_{T_{2},\bar{\epsilon}}^{\vec{f}}) & =\\
 & =f((L_{T_{1},\bar{\epsilon}}^{\vec{f}}\cap R_{T_{2},\bar{\epsilon}}^{\vec{f}}\cap L_{T,\bar{\epsilon}}^{\vec{f}})\cup(L_{T_{1},\bar{\epsilon}}^{\vec{f}}\cap R_{T_{2},\bar{\epsilon}}^{\vec{f}}\cap(L_{T,\bar{\epsilon}}^{\vec{f}})^{c}))\\
 & =f(L_{T_{1},\bar{\epsilon}}^{\vec{f}}\cap R_{T_{2},\bar{\epsilon}}^{\vec{f}}\cap L_{T,\bar{\epsilon}}^{\vec{f}})+f(L_{T_{1},\bar{\epsilon}}^{\vec{f}}\cap R_{T_{2},\bar{\epsilon}}^{\vec{f}}\cap R_{T,\bar{\epsilon}}^{\vec{f}})+f(L_{T_{1},\bar{\epsilon}}^{\vec{f}}\cap R_{T_{2},\bar{\epsilon}}^{\vec{f}}\cap\{T=\bar{\epsilon}\}).
\end{array}
\]
\noindent Thus, if $f(L_{T_{1},\bar{\epsilon}}^{\vec{f}}\cap R_{T_{2},\bar{\epsilon}}^{\vec{f}}\cap L_{T,\bar{\epsilon}}^{\vec{f}})>0,$
then $f(R_{T_{2},\bar{\epsilon}}^{\vec{f}}\cap L_{T,\bar{\epsilon}}^{\vec{f}})>0,$
which contradicts the second condition of \eqref{eq: Transitivity condition};
otherwise if $f(L_{T_{1},\bar{\epsilon}}^{\vec{f}}\cap R_{T_{2},\bar{\epsilon}}^{\vec{f}}\cap R_{T,\bar{\epsilon}}^{\vec{f}})>0$,
then $f(L_{T_{1},\bar{\epsilon}}^{\vec{f}}\cap R_{T,\bar{\epsilon}}^{\vec{f}})>0,$
which contradicts the first condition of \eqref{eq: Transitivity condition}.
Otherwise, $f(L_{T_{1},\bar{\epsilon}}^{\vec{f}}\cap R_{T_{2},\bar{\epsilon}}^{\vec{f}}\cap\{T=\bar{\epsilon}\})>0$
implies that $f(L_{T_{1},\bar{\epsilon}}^{\vec{f}}\cap\{T=\bar{\epsilon}\})>0.$
In addition, since $\{L_{T_{1},\bar{\epsilon}+\frac{1}{n}}^{\vec{f}}\}_{n>\lceil\frac{1}{1-\bar{\epsilon}}\rceil}$
is increasing to $L_{T_{1},\bar{\epsilon}}^{\vec{f}}$ and $R_{T,\bar{\epsilon}}^{\vec{f}}\subset\{R_{T,\bar{\epsilon}+\frac{1}{n}}^{\vec{f}}\}$
for all $n$ it follows that there exists a sufficiently large $n'$
and $\epsilon'=\bar{\epsilon}+\frac{1}{n'}$ such that $f(L_{T_{1},\epsilon'}^{\vec{f}}\cap R_{T,\epsilon'}^{\vec{f}})>0,$
which again contradicts the first condition of \eqref{eq: Transitivity condition}.

Case 2: $f(L_{T_{2},\bar{\epsilon}}^{\vec{f}}\cap R_{T_{1},\bar{\epsilon}}^{\vec{f}})>0.$
The contradiction follows analogously from Case 1 and hence omitted.\Halmos
\endproof
\vspace{0cm}

\begin{claim}
\label{Claim, if T reasonable then P1(A intersect AT0)}If $T$ is
reasonable then for all $\vec{f}$ and $\epsilon\in(\frac{1}{2},1)$
and for all measurable set $A$
\[
f(A\cap R_{T,\epsilon}^{\vec{f}})>0\implies g(A\cap R_{T,\epsilon}^{\vec{f}})>0.
\]
\noindent (similarly for $g$ where $\epsilon\in(0,\frac{1}{2})$).
\end{claim}
\proof{Proof.}\
Let $\vec{f}$ and $\epsilon\in(\frac{1}{2},1)$ and a measurable
set $A$, and (w.l.o.g.) assume by contradiction that
\[
f(A\cap R_{T,\epsilon}^{\vec{f}})>0\implies g(A\cap R_{T,\epsilon}^{\vec{f}})=0.
\]
\noindent Since $0=g(A\cap R_{T,\epsilon}^{\vec{f}})<(\frac{1-\epsilon}{\epsilon})f(A\cap R_{T,\epsilon}^{\vec{f}})$
and $T$ is reasonable \eqref{eq:reasonableness condition} yields
that $f(A\cap R_{T,\epsilon}^{\vec{f}}\cap L_{T,\epsilon}^{\vec{f}})>0,$
which contradicts the fact as $R_{T,\epsilon}^{\vec{f}},L_{T,\epsilon}^{\vec{f}}$
are disjoint sets.\Halmos
\endproof
\vspace{0cm}

\proof{\textbf{Proof of Lemma \ref{lem:reasonable implies m.s}}.}
W.l.o.g. let $A$ be such that: $f(A)=1,\ g(A)=0,$ and let $\epsilon\in(\frac{1}{2},1)$.
Assume that $f(A\cap R_{T,\epsilon}^{\vec{f}})>0,$ $T$ is reasonable,
therefore applying Claim \eqref{Claim, if T reasonable then P1(A intersect AT0)}
with the set $A$ yields
\[
g(A\cap R_{T,\epsilon}^{\vec{f}})>0
\]
which contradicts the assumption that $g(A)=0.$ Hence, $f(A\cap R_{T,\epsilon}^{\vec{f}})=0.$
On the other hand, since the left-hand side of condition \eqref{eq:reasonableness condition}
is satisfied trivially for $A,$ we are provided with $f(A\cap L_{T,\epsilon}^{\vec{f}})\geq f(A\cap L_{T,\epsilon}^{\vec{f}})>0.$
As a result,
\[
1=f(A)=f(A\cap R_{T,\epsilon}^{\vec{f}})+f(A\cap L_{T,\epsilon}^{\vec{f}})+f(A\cap\{T=\epsilon\})
\]
\noindent and therefore $f(A\cap L_{T,\epsilon^{1.01}}^{\vec{f}})=1.$
Similarly, assuming that $f(B)=0,\ g(B)=1$ we obtain that $g(R_{T,\epsilon^{0.99}}^{\vec{f}})=1$
for all $\epsilon\in(0,\frac{1}{2}).$ Since $A_{T,f}^{\vec{f}}\subset L_{T,\epsilon}^{\vec{f}}$
for all $\epsilon\in(\frac{1}{2},1)$ and $L_{T,\epsilon}^{\vec{f}}$
is decreasing as $\epsilon\rightarrow1$ (as the partition is refined)
it follows that $f(A_{T,f}^{\vec{f}})=f(\underset{\epsilon}{\bigcap}L_{T,\epsilon}^{\vec{f}})=1$
and the result follows.\Halmos
\endproof
\vspace{0cm}

\setcounter{equation}{0} \renewcommand{\theequation}{B.1.\arabic{equation}}
\subsection{{Decisiveness in finite time}}\

\vspace{0cm}

The proof of Theorem \ref{Th: second main theorem (F=000026L)} is
generalized to the case where the number of elements, $|\Omega|$,
is arbitrary and it is relied on achieving a uniform bound on the
up-crossing probability of any non-negative supermartingale which
admits sufficiently (finite) many fluctuations. 

\proof{\textbf{Proof of Theorem \ref{Th: second main theorem (F=000026L)}.}}
Let $\epsilon\in(0,1)$. We will show that there exists a uniform
constant $K=K(\epsilon)$ such that on the set of histories $\omega^{t}$
of $f-probability-(1-\epsilon)$, and for all $n>0,$ only two scenarios
are possible; if there exists a subsequence of times $(t_{i})_{i=1}^{K+1}\subset\{1,...,n\}$,
and there exists a subsequence of corresponding outcomes $(\varpi_{t_{i}})_{i=1}^{K+1}\subset\Omega^{K+1}$
such that $|f((\omega,\vec{f})^{t_{i}-1})[\varpi_{t_{i}}]-g((\omega,\vec{f})^{t_{i}-1})[\varpi_{t_{i}}]|\geq\epsilon$
for all $1\leq i\leq K+1$, then, the limit of ${\cal D}$ is strictly
greater than $(1-\epsilon)$, and more importantly, the value of ${\cal D}$
at time $n$, ${\cal D}_{n},$ is $\epsilon-close$ for all ranks
from time $n$ onward. In all other scenarios, $|f((\omega,\vec{f})^{t-1})[\varpi]-g((\omega,\vec{f})^{t-1})[\varpi]|<\epsilon$
for all $\varpi\in\Omega$ in all but $K$ periods $t$ in $\{1,...,n\}$. 

\paragraph*{Construction of the faster process}

As in \citet{Fudenberg-Levine-1992}, define an increasing sequence
of stopping times $\{\tau_{k}\}_{k=0}^{\infty}$ relative to $\{D_{f}^{t}g\}$
and $\epsilon$ inductively as follows. First set $\tau_{0}=0$ and
if $\tau_{k-1}(\omega)=\infty$ set $\tau_{k}(\omega)=\infty.$ If$\tau_{k-1}(\omega)<\infty$
set $\tau_{k}(\omega)$ to be the smallest integer $t>\tau_{k-1}(\omega)$
such that either 
\begin{equation}
f(\omega^{t-1})>0\ and\ f(\{\bar{\omega}\in\Omega^{\infty}:\ |\text{\ensuremath{\frac{\text{\ensuremath{D_{f}^{t}g}}(\bar{\omega})}{\text{\ensuremath{D_{f}^{t-1}g}}(\bar{\omega}))}}-1|>\ensuremath{\frac{\epsilon}{|\Omega|}}\}}|\ \omega^{t-1}\})>\frac{\epsilon}{|\Omega|}\label{eq:condition 1}
\end{equation}
\noindent or
\begin{equation}
\frac{D_{f}^{t}g(\omega)}{D_{f}^{\tau_{k-1}}g(\omega)}-1\geq\frac{\epsilon}{2|\Omega|}.\label{eq:condition 2}
\end{equation}
\noindent If there is no such $t,$ set $\tau_{k}(\omega)=\infty.$
Now define the process $\{\tilde{{\cal D}}_{k}\}_{k=0}^{\infty}$
by $\tilde{{\cal D}}_{k}=D_{f}^{\tau_{k}}g$ if $\tau_{k}<\infty$
and $\tilde{{\cal D}}_{k}=0$ if $\tau_{k}=\infty.$
Now, From \citet{Fudenberg-Levine-1992}, Lemma 4.1, $(D_{f}^{t}g(\omega)$
$:=\frac{g(\omega^{t})}{f(\omega^{t})})_{t>0}$ is a supermartingale;
hence from a standard result, the process $\{\tilde{{\cal D}}_{k}\}_{k=0}^{\infty}$
is a supermartingale. Furthermore, by \citet{Fudenberg-Levine-1992},
Lemma 4.3, $\{\tilde{{\cal D}}_{k}\}_{k=0}^{\infty}$ is an active
supermartingale with activity $\frac{\epsilon}{2|\Omega|}.$

Applying Theorem \ref{Th A.1 (F=000026L)}  with $\epsilon,$ $|\Omega|,$
$acitivity=\frac{\epsilon}{2|\Omega|},$ and $\tilde{{\cal D}}_{0}\equiv1,$
there exists an integer $K=K(\epsilon)>0$ (depending only on these
variables) such that for any active supermartingale $\{\tilde{{\cal D}}_{k}\}$
with activity $\frac{\epsilon}{2|\Omega|},$ one has
\begin{equation}
f(\underset{k>K}{sup}\tilde{{\cal D}}_{k}<\epsilon)>1-\epsilon.\label{eq:supL-1}
\end{equation}
In addition, by \citet{Fudenberg-Levine-1992}, Lemma 4.2, if $|f((\omega,\vec{f})^{t})[\varpi]-g((\omega,\vec{f})^{t})[\varpi]|>\epsilon,$
for some $\varpi\in\Omega$ then condition \eqref{eq:condition 1}
holds. Consequently, the process $\{\tilde{{\cal D}}_{k}\}_{k=0}^{\infty}$
takes into account all observations where $|f((\omega,\vec{f})^{t})[\varpi]-g((\omega\vec{f})^{t})[\varpi]|>\epsilon$
for some $\varpi\in\Omega$ and omits only observations where $|f((\omega,\vec{f})^{t})[\varpi]-g((\omega,\vec{f})^{t})[\varpi]|\leq\epsilon$
for all $\varpi\in\Omega$ (although, by condition \eqref{eq:condition 2},
not necessarily all of them). 

As a result, under the assumption that expert $f$ is truthful (meaning,
the realizations are generated via $f$), there exists a constant
$K=K(\epsilon)$, which does not depend on the true process $f$ or
the forecasting strategy $g,$ so that on the set of histories, $\omega^{t},$
of probability $(1-\epsilon)$ under $f$, in all but $K$ periods
either $|f((\omega,\vec{f})^{t})[\varpi]-g((\omega,\vec{f})^{t})[\varpi]|\leq\epsilon$
for all $\varpi\in\Omega$ or $D_{f}^{t}g(\omega)<\epsilon.$ 

Now assume that there exist $\text{\ensuremath{K+1} periods }(t_{i})_{i=1}^{K+1}\subset\{1,...,n\}$
and $(\varpi_{t_{i}})_{i=1}^{K+1}\subset\Omega^{K+1}$ such that $|f((\omega,\vec{f})^{t_{i}-1})[\varpi_{t_{i}}]-g((\omega,\vec{f})^{t_{i}-1})[\varpi_{t_{i}}]|\geq\epsilon$
for all $1\leq i\leq K+1$ with $f(\omega^{t_{i}-1})>0$ and let $n>K+1.$
Then equation \eqref{eq:supL-1} ensures us that with $f-probability-(1-\epsilon)$ 

\begin{equation}
\tilde{{\cal D}}_{K+1}=D_{f}^{\tau_{K+1}}g<\epsilon\label{eq:9}
\end{equation}

\noindent where by condition \eqref{eq:condition 2} for any $t\geq n\geq\tau_{K+1}$
we obtain that either $\tilde{{\cal D}}_{t}$ drops below $\epsilon$
or 

\begin{equation}
D_{f}^{t}g(\omega)<D_{f}^{\tau_{K+1}}g(\omega)(1+\frac{\epsilon}{2})<\epsilon(1+\epsilon)\label{eq:10}
\end{equation}

\noindent and hence it cannot exceed $\epsilon(1+\epsilon$). 

We conclude that there exists a constant $K$, which does not depend
on the forecasting strategies $f,g$, such that for any sufficiently
large $n>K$, with $f$ - probability - $(1-\epsilon);$ if there
exist $K+1$ periods in which $f$ and $g$ are slightly different
above $\epsilon$ along a play path then the likelihood ratio at any
point $t$ after $n$ never exceeds $\epsilon(1+\epsilon).$ 

\vspace{0cm}

\noindent Now from \eqref{eq:9} and \eqref{eq:10} we conclude that
with $f$ - probability - $(1-\epsilon)$, either $|f((\omega,\vec{f})^{t-1})[\varpi]-g((\omega,\vec{f})^{t-1})[\varpi]|<\epsilon$
for all $\varpi\in\Omega$ in all but $K$ periods $t$ in $\{1,...,n\}$,
or

\begin{equation}
1-\frac{\epsilon(1+\epsilon)}{1+\epsilon(1+\epsilon)}=\frac{1}{1+\epsilon(1+\epsilon)}<\frac{1}{1+D_{f}^{t}g(\omega)}={\cal D}_{t}(\omega,\vec{f})\leq1\label{eq:11}
\end{equation}

\noindent for all $t\geq n,$ and as a result, the liminf of ${\cal D}$
is always greater than $1-\frac{\epsilon(1+\epsilon)}{1+\epsilon(1+\epsilon)}.$
Consequently, for all $t\geq n$ inequality \eqref{eq:11} yields
\[
1-\frac{\epsilon(1+\epsilon)}{1+\epsilon(1+\epsilon)}<|{\cal D}_{n}(\omega,\vec{f})-{\cal D}_{t}(\omega,\vec{f})|\leq1,
\]
\noindent and hence $|{\cal D}_{n}(\omega,\vec{f})-{\cal D}_{t}(\omega,\vec{f})|<\frac{\epsilon(1+\epsilon)}{1+\epsilon(1+\epsilon)}$,
which, together with the first scenario, holds with $f$ - probability
- $(1-\epsilon)$. Since $\frac{\epsilon(1+\epsilon)}{1+\epsilon(1+\epsilon)}<\epsilon$,
the result follows.\Halmos 
\endproof
\vspace{0cm}

\section{{\label{sec:Appendix B The-Cross-Calibration}The cross-calibration test}}\

We now restate the cross-calibration test as suggested by \citet*{Feinberg-Stewart-2008}.
Fix a positive integer $N>4$ and divide the interval $[0,1]$ into
$N$ equal closed subintervals $I_{1},...,I_{N},$ so that $I_{j}=[\frac{j-1}{N},\frac{j}{N}],\ 1\leq j\leq N$.
All results in their paper hold when $[0,1]$ is replaced with the
set of distributions over any finite set $\Omega$ and the intervals
$I_{j}$ are replaced with a cover of the set of distributions by
sufficiently small closed convex subsets. At the beginning of each
period $t=1,2...$ , all forecasters (or experts) $i\in\{0,..,M-1\}$
simultaneously announce predictions $I_{t}^{i}\in\{I_{1},...,I_{N}\}$,
which are interpreted as probabilities with which the realization
$1$ will occur in that period. We assume that forecasters observe
both the realized outcome and the predictions of the other forecasters
at the end of each period. 

The cross-calibration test is defined over outcomes $(\omega_{t},I_{t}^{0},...,I_{t}^{M-1})_{t=1}^{\infty}$,
which specify, for each period $t$, the realization $\omega_{t}\in\Omega$,
together with the prediction intervals announced by each of the $M$
forecasters. Given any such outcome and any $M$ - tuple $l=(I_{l^{0}},...,I_{l^{M-1}})\in\{I_{1},...,I_{N}\}^{M}$,
let $\zeta_{t}^{l}=1_{I_{t}^{i}=I_{l^{i}},\ \forall i=0,...,M-1},$
and
\[
\nu_{t}^{l}=\stackrel[n=1]{t}{\sum}\zeta_{n}^{l},
\]
\noindent which represents the number of times that the forecast profile
$l$ is chosen up to time $n$. For $\nu_{t}^{l}>0$, the frequency
$f_{n}^{l}$ of realizations conditional on this forecast profile
is given by
\[
f_{t}^{l}=\frac{1}{\nu_{t}^{l}}\stackrel[n=1]{t}{\sum}\zeta_{n}^{l}\omega_{n}.
\]
Forecaster $i$ passes the cross-calibration test at the outcome $(\omega_{t},I_{t}^{0},...,I_{t}^{M-1})_{t=1}^{\infty}$
if 
\[
\underset{t\rightarrow\infty}{limsup}|f_{t}^{l}-\frac{2l^{i}-1}{2N}|\leq\frac{1}{2N}
\]
\noindent for every $l$ satisfying $\underset{t\rightarrow\infty}{lim}\nu_{t}^{l}=\infty$. 

In the case of a single forecaster, the cross-calibration test reduces
to the classic calibration test, which checks the frequency of realizations
conditional on each forecast that is made infinitely often. With multiple
forecasters, the cross-calibration test checks the empirical frequencies
of the realization conditional on each profile of forecasts that occurs
infinitely often. Note that if an expert is cross-calibrated, he will
also be calibrated.

\end{APPENDICES}

% Appendix here
% Options are (1) APPENDIX (with or without general title) or 
%             (2) APPENDICES (if it has more than one unrelated sections)
% Outcomment the appropriate case if necessary
%
% \begin{APPENDIX}{<Title of the Appendix>}
% \end{APPENDIX}
%
%   or 
%
\def\notesname{Endnotes\\}
\theendnotes

%\section*{{\normalsize{}Endnotes}}

% \begin{APPENDICES}
% \section{<Title of Section A>}
% \section{<Title of Section B>}
% etc
% \end{APPENDICES}

% Acknowledgments here
%\section*{Acknowledgments}
%Smorodinsky gratefully acknowledges United States-Israel Binational
%Science Foundation and the German-Israel Foundation grant I-1419-118.4/2017,
%the Ministry of Science and Technology grant 19400214, Technion VPR
%grants, and the Bernard M. Gordon Center for Systems Engineering at
%the Technion.

\vspace{0cm}

%\bibliographystyle{plainnat}
%\bibliography{A_cardinal_comparison_of_experts}

\vspace{0cm}
% Enter the text of acknowledgments here

% References here (outcomment the appropriate case) 

% CASE 1: BiBTeX used to constantly update the references 
%   (while the paper is being written).
\bibliographystyle{informs2014} % outcomment this and next line in Case 1
%\bibliography{<your bib file(s)>} % if more than one, comma separated
%\bibliography{MOR-template}

\bibliography{A_cardinal_comparison_of_experts}

% CASE 2: BiBTeX used to generate mypaper.bbl (to be further fine tuned)
%\input{mypaper.bbl} % outcomment this line in Case 2

\end{document}